\documentclass[12pt,a4paper,preprint,superscriptaddress]{revtex4-1}
\usepackage{float}
\usepackage{amsmath,amssymb}
\usepackage[utf8]{inputenc}
\usepackage{graphicx}
\usepackage[english]{babel}

\usepackage{xcolor}

\makeatletter\AtBeginDocument{\let\@elt\relax}\makeatother

\begin{document}

\title{Physics-separating artificial neural networks for predicting initial stages of Al sputtering and thin film deposition in Ar plasma discharges}
\date{\today}
\author{Tobias Gergs}
\email[]{tobias.gergs@rub.de}
\author{Thomas Mussenbrock}
\email[]{thomas.mussenbrock@rub.de}
\affiliation{Chair of Applied Electrodynamics and Plasma Technology, Department of Electrical Engineering and Information Science, Ruhr University Bochum, 44780 Bochum, Germany}
\author{Jan Trieschmann}
\email[]{jt@tf.uni-kiel.de}
\affiliation{Theoretical Electrical Engineering, Department of Electrical and Information Engineering, Kiel University, Kaiserstraße 2, 24143 Kiel, Germany}
\affiliation{Kiel Nano, Surface and Interface Science KiNSIS, Kiel University, Christian-Albrechts-Platz 4, 24118 Kiel, Germany}

\begin{abstract}
Simulations of Al thin film sputter depositions rely on accurate plasma and surface interaction models. Establishing the latter commonly requires a higher level of abstraction and means to dismiss the fundamental atomic fidelity. Previous works on sputtering processes addressed this issue by establishing machine learning surrogate models, which include a basic surface state (i.e., stoichiometry) as static input. In this work, an evolving surface state and defect structure are introduced to jointly describe sputtering and growth with physics-separating artificial neural networks. The data describing the plasma-surface interactions stem from hybrid reactive molecular dynamics/time-stamped force bias Monte Carlo simulations of Al neutrals and Ar$^+$ ions impinging onto Al(001) surfaces. It is demonstrated that the fundamental processes are comprehensively described by taking the surface state as well as defect structure into account. Hence, a machine learning plasma-surface interaction surrogate model is established that resolves the inherent kinetics with high physical fidelity. The resulting model is not restricted to input from modeling and simulation, but may similarly be applied to experimental input data.
\end{abstract}

\maketitle

\newpage

%%%%%%%%%%%%%%%%%%%%%%%%%%%%%%%%%%%%%%%%%%%%%%%%%%%

\section{Introduction}
\label{sec:introduction}

Physical vapor deposition is commonly used for the sputter deposition of thin films. Various processes and types of gas discharges may be exploited, whereas a direct current or radio frequency voltage of a few 100\,V to kV are typically used to sustain a plasma. Inert or reactive gases may be used depending on the application and the desired thin film. Sputtering processes have in common that the strong electric field in the plasma sheaths accelerates ions (e.g., Ar$^+$) towards the electrodes. \cite{kelly_magnetron_2000,gudmundsson_physics_2020,baptista_sputtering_2018,rossnagel_handbook_1990}
At the target surface, the ion bombardment leads to a collision cascade and eventually emission of material. This is referred to as sputtering and be can be studied theoretically by performing transport of ions in matter (TRIM), Monte Carlo (MC), or molecular dynamics (MD) based simulations. After release of the sputtered atoms from the surface with the characteristic Sigmund-Thompson distribution \cite{biersack_monte_1980, eckstein_sputtering_1984, moller_tridyn_1984, voter_introduction_2007, graves_molecular_2009, neyts_molecular_2017,sigmund_theory_1969,sigmund_theory_1969-1,thompson_ii_1968},  they are subject to collisional transport through the plasma \cite{trieschmann_transport_2015,gudmundsson_physics_2020,rossnagel_deposition_1988,somekh_thermalization_1984}. Collisions with electrons may ionize the otherwise neutral atoms. Transport phenomena can be resolved by either fluid or particle in cell/Monte Carlo collision (PIC/MCC) models coupled to test-particle methods (TPM) or direct simulation Monte Carlo (DSMC) models \cite{birdsall_plasma_1991, dijk_plasma_2009, serikov_particle--cell_1999, somekh_thermalization_1984, turner_monte_1989, trieschmann_transport_2015}. When the sputtered atoms arrive at the opposing substrate surface, they contribute to thin film growth. This kind of surface interaction can be similarly simulated by effective phenomenological, MC, or MD models \cite{voter_introduction_2007, graves_molecular_2009, neyts_molecular_2017,depla_reactive_2008}. 

The different length and time scales of the solid and the gas/plasma states of matter render a joint simulation framework with preserved atomic fidelity impractical  \cite{kruger_machine_2019, bird_molecular_1994, lieberman_principles_2005, callister_materials_2013}. Hence, surface and plasma simulations need to be evaluated separately, whereas often a priori evaluated particle fluxes onto the surfaces and effective surface kinetics are considered (e.g. Berg model \cite{berg_fundamental_2005, depla_reactive_2008}, Sigmund-Thompson theory \cite{thompson_ii_1968, sigmund_theory_1969, sigmund_theory_1969-1}), look-up tables).

This issue has been addressed in previous works by proposing data-driven machine learning plasma-surface interaction (PSI) surrogate models that are intended to complement plasma simulations \cite{kruger_machine_2019,gergs_efficient_2022,diaw_multiscale_2020, ulissi_address_2017, kino_characterization_2021, adamovich_2022_2022, anirudh_2022_2022, preuss_bayesian_2019}. In particular, a multi layer perceptron (MLP) was trained to generalize on data describing the sputtering of a Ti$_{0.5}$Al$_{0.5}$ composite target, simulated with TRIDYN \cite{kruger_machine_2019,moller_tridyn_1984}. TRIDYN was also used in a subsequent study to generate data for Ti$_{1-x}$Al$_x$ composite targets. The stoichiometry $x$ was introduced to the model as an additional degree of freedom and preliminary surface state. Moreover, the MLP was replaced by a $\beta$-variational autoencoder establishing a reduced latent space representations of the sputtered particle energy angular distribution functions as well as a dedicated mapper network correlating them with the input (i.e., Ar$^+$ ion energy distribution functions) \cite{gergs_efficient_2022,moller_tridyn_1984,kingma_auto-encoding_2013, rezende_stochastic_2014}. 

In this work, the concept of a machine learning surface surrogate model is advanced by -- among other aspects -- exploiting hybrid reactive molecular dynamics/time-stamped force-biased Monte Carlo (RMD/tfMC) simulation data of sputtering and deposition of Al thin films with Ar working gas \cite{bal_time_2014,mees_uniform-acceptance_2012,neyts_combining_2014,gergs_molecular_2022}. The manuscript is structured as follows: In Section~\ref{ssec:setup} the setup depicting the surface interactions is introduced. In Section~\ref{sec:methods}, the utilized methods and the corresponding parameters are detailed. The results are collected and discussed in Section~\ref{sec:results}. Finally, conclusions are drawn in Section~\ref{sec:conclusion}.

%%%%%%%%%%%%%%%%%%%%%%%%%%%%%%%%%%%%%%%%%%%%%%%%%%%

\section{Setup}
\label{ssec:setup}

The considered plasma-surface interactions relate to a previous simulation study investigating the sputter deposition of Al thin films with Ar working gas, and specifically Ar$^+$ ion and Al neutral impingement on an Al(100) surface. The evaluation procedure and physical interpretation is detailed in  \cite{gergs_molecular_2022}. In summary, hybrid reactive molecular dynamics/time-stamped force-biased Monte Carlo (RMD/tfMC) simulations have been conducted to describe individual surface interactions \cite{bal_time_2014,mees_uniform-acceptance_2012,neyts_combining_2014}. The simulations have been performed using the open-source Large-scale Atomic/Molecular Massively Parallel Simulator (LAMMPS) \cite{plimpton_fast_1995,thompson_lammps_2022}. The crucial role of Ar atoms incorporated in the Al surface has been identified for the description of the properties of the deposited thin film (e.g., stress, defect structure). A rigorous treatment of the incorporated Ar atoms has been found to be of particular relevance in this regard. In the following, the key aspects of the simulation setup and the provided data are reviewed:

\paragraph*{Particle fluxes to the surface}
The particle fluxes incident from the plasma to the surfaces are defined as follows. Two particle doses consecutively impinge onto an Al(001) surface, each of $1.21\times10^{15}$\,particles/cm$^2$. The fluxes consist of five combinations of Al neutral to Ar$^+$ ion flux ratios $\Gamma_{\mathrm{Al}}^\mathrm{in}/\Gamma_{\mathrm{Ar^+}}^\mathrm{in}\in[0,1]$. 20 different Ar$^+$ ion energies $E_{\mathrm{Ar^+}}^\mathrm{in}$ are selected from a square root energy axis in the range of 3 to 300 eV. Al neutrals are assumed to be thermalized ($k_\mathrm{b}T=7$\,eV) \cite{gergs_molecular_2022}. The particle fluxes from the plasma to the surface are characterized by the descriptors $P=\{E_{\mathrm{Ar^+}},\Gamma_{\mathrm{Al}}^\mathrm{in}/\Gamma_{\mathrm{Ar^+}}^\mathrm{in}\}$.

\paragraph*{Surface defect structure}
The ongoing ion bombardment leads to surface reconstructions due to the accompanying collision cascades and due to point defects (e.g., vacancies, interstitials). The latter result from sputtering (emission) as well as forward sputtering (peening). The defect structure after the 1st and 2nd dose of incident particles is quantified by a ring statistical connectivity profile \cite{drabold2005models,cobb1996ab,zhang2000structural}. It describes the systems' bond topology and consists of four properties: i) The number of rings per supercell divided by the number of nodes $R_{\mathrm{C},r}$ for rings with $r$ nodes; ii) The fraction of nodes at the origin of at least one ring with $r$ nodes $P_{\mathrm{N},r}$; iii) The fraction of nodes with $r$ sized rings as shortest closed paths $P_{\mathrm{min},r}$; iv) The fraction of nodes with $r$ sized rings as longest closed paths $P_{\mathrm{max},r}$ \citep{drabold2005models,cobb1996ab,zhang2000structural,le2010ring,gergs_molecular_2022}. The defect structure is characterized by the descriptors $D=\{R_{\mathrm{C},r},P_{\mathrm{N},r},P_{\mathrm{min},r},P_{\mathrm{max},r}\}$ for $r\in [3,6]$.

\paragraph*{Surface state and properties}
Defects are assumed to be the predominant cause for changes of the surface state (e.g., film stress, mechanical properties) during the ion bombardment \cite{karimi_aghda_unravelling_2021}. Hence, it is assumed that the defect structure $D$ and the surface state descriptors $S_\mathrm{s}=\{$mass density $\rho$, Ar concentration $x_\mathrm{Ar}$, biaxial stress $\frac{\sigma_{xx}+\sigma_{yy}}{2}$, shear stress $\tau_{xy}$\} equivalently describe the system state of the surface. A hypothesis that will be revisited later by trying to derive one from another. The surface state is further characterized by an additional set of surface properties, $A_\mathrm{s}=\{$Ar gas porosity $\phi_\mathrm{Ar}$, Al vacancy density $N_\mathrm{v_{Al}}$, root mean square roughness $R_\mathrm{RMS}$\} \cite{gergs_molecular_2022}.

\paragraph*{Particle fluxes from the surface}
Finally, the fluxes of particles emitted from the surface due to reflection or sputtering are characterized by their outgoing energy distribution functions (EDFs) $f$. These are evaluated by collecting the individual particle trajectories. The details of how the outgoing particle information is processes are presented in Section~\ref{ssec:data}.

%%%%%%%%%%%%%%%%%%%%%%%%%%%%%%%%%%%%%%%%%%%%%%%%%%%

\section{Methods}
\label{sec:methods}

In the following we proceed with an extension of the strategies detailed in previous publications on machine learning surface surrogate models \cite{kruger_machine_2019,gergs_efficient_2022}. Specifically, after assembling the data set from hybrid RMD/tfMC simulations and defining the relevant descriptors, the data is pre-processed in terms of its data structure and normalization. In this work, data augmentation is added thereafter, as elaborated below. The data set is divided for training, validation, and testing and a physics-separating artificial neural network (ANN) is optimized and trained subsequently. Finally, the model may be assessed given a number of metrics specified below.

\subsection{Data preparation, training and metrics}
\label{ssec:data}

\paragraph{Outgoing EDFs pre-processing}
The ensemble of outgoing particles are characterized by the individual particle properties (especially kinetic energies). Equivalently they can be depicted by their respective EDFs. The EDFs are obtained from phase-space information of the individual particle in two steps. (1) A kernel density estimator (KDE) with tophat kernel is used to obtain a continuous representation of the EDFs based on the particle energies. KDE is favored instead of simple histogram binning providing a more accurate description despite the statistical challenges. Specifically, the simulation case with the best statistical description, that is the largest number of emitted (i.e., reflected, sputtered) particles, is used to fit the KDE kernel parameters. The kernel bandwidth is determined by optimizing the 10-fold cross validated log-likelihood, giving 5.01\,eV for Al and 2.75\,eV for Ar. The probability densities obtained with the KDE are normalized by enforcing the integral over energy to equal the number of emitted particles divided by the number of incident Ar$^+$ ions. Hence, information about the reflection ratio and sputtering yields are included. (2) The KDE is subsequently applied to sample discrete EDFs on a regular grid in the range of 0\,eV to 20\,eV with a step size of 0.5\,eV for both species Al and Ar.

\paragraph{Data normalization}
All data (input data $x$ and output data $y$) is normalized to range from zero to one: $x\rightarrow (x-x_\mathrm{min})/(x_\mathrm{max}-x_\mathrm{min}+\epsilon)$ and $y\rightarrow (y-y_\mathrm{min})/(y_\mathrm{max}-y_\mathrm{min}+\epsilon)$. The min/max values are extracted from the training data set and stored for the final denormalization. For the EDFs, the maximum value per species is used as a reference. Notably, this normalization is merely relevant for an optimal training procedure.

\paragraph{Data set size and splitting}
In total $n_\mathrm{data}=200$ individual cases represent respective training samples from a combination of $2\times5\times20$ varied parameters. The data set is shuffled and split into 80 \% training, 10 \% validation, and 10 \% test set (i.e., $n_\mathrm{data}^\mathrm{train}=160$, $n_\mathrm{data}^\mathrm{val}=20$, $n_\mathrm{data}^\mathrm{test}=20$) for a study of the hyperparameters (HPs) of the ANN and its subsequent training.

\paragraph{Data augmentation}
A virtual (artificial) extension of such a small data set by augmentation with ``combined'' data samples improves the training robustness in spite of providing no further information (but only a hypothesis) to the network \cite{oviedo_fast_2019,li_genetic_2014,choudhury_artificial_2011}. In this work, a constrained version of mixup augmentation is proposed. Mixup augmentation is commonly used for image classification, but in this work found to facilitate training a more robust regression model \cite{zhang_mixup_2018}. Augmented input $\hat{x} = \lambda x_i + (1-\lambda) x_j$ and corresponding output samples $\hat{y} = \lambda y_i + (1-\lambda) y_j$ are computed for samples $i$ and $j$, respectively. The factor $\lambda$ is sampled from the beta distribution $\lambda\sim\text{Beta}(\alpha,\beta)$, whereas $\alpha=\beta$ is considered as HP. $\hat{x}$ and $\hat{y}$ are recomputed each epoch \cite{zhang_mixup_2018}. However, sample pairs $i$ and $j$ are only mixed when they fulfill the constraint $\sqrt{(k_i-k_j)^2+(l_i-l_j)^2+(m_i-m_j)^2}<r_\mathrm{c}$. $k$, $l$, and $m$ specify the indices along the incident Ar$^+$ ion energy axis $E_{\mathrm{Ar^+},k}^\mathrm{in}$, the Al neutral to Ar$^+$ ion flux ratio axis $\Gamma_{\mathrm{Al},l}^\mathrm{in}/\Gamma_{\mathrm{Ar^+},l}^\mathrm{in}$, and the 1st dose ($m=0$) or the 2nd dose ($m=1$) of impinging particles, respectively. The augmentation range $r_\mathrm{c}$ is considered as HP. It allows for a continuous transition from no augmentation (i.e., $r_\mathrm{c}\rightarrow0$) to ordinary mixup augmentation (i.e., $r_\mathrm{c}\rightarrow\infty$). Mixing of the data can be interpreted as providing the hypothesis of a linear superposition of sample pairs with a maximum index radius $r_\mathrm{c}$ to the network. This mixup combination is at most conducted once per sample pair. The training set is then replaced by all combinations of mixup training set sample pairs fulfilling this constraint. This means that for instance for $r_\mathrm{c}\rightarrow\infty$ the number of augmented training set samples is squared (i.e., $n_\mathrm{data,aug}^\mathrm{train} = {n_\mathrm{data}^\mathrm{train}}^2$).

\paragraph{Training procedure and metrics}
The mean absolute errors (MAEs) of the model's predictions on each reference training batch are backpropagated through the network to update the network weights (degrees of freedom). The stochastic gradient descent algorithm adaptive moment estimation (Adam) is applied \cite{kingma_adam_2015}. The batch size $n_\mathrm{batch}$ is chosen to be as close as possible to equal 32 while fulfilling $|n_\mathrm{batch}-n_\mathrm{data,aug}^\mathrm{train}\%n_\mathrm{batch}|\leq4$. This requirement ensures that all data samples contribute almost equally to the training progress.  The coefficient of determination $R^2$ is utilized as an additional metric to assess the network's prediction accuracy. A coefficient of determination equal to one resembles a fully explained residual variance.

%purpose of val set (training)
The learning rate is set to $10^{-3}$ for an initial simulated annealing phase (described in Section~\ref{ssec:model}). Subsequently, it is reduced by a factor of 10 whenever the validation loss is not smaller than its previous minimum value over the course of 10 epochs. Early stopping ends the training phase when there is no further reduction of the validation loss in 25 epochs.

\paragraph{Hyperparameter study}
After training, the validation loss may be used to find a suitable set of HPs as detailed in Section~\ref{ssec:hpStudy}. 10-fold Monte Carlo cross validation (MCCV) is applied to address the data splitting induced variance (potential bias) in model training and validation, as well as its test error in prediction. The last is used to asses the model's performance on unknown data (no feedback to the model). Once the final set of HPs is selected, the corresponding network is trained with the validation and augmented training set by performing a 100-fold MCCV. To obtain final training metrics, the MAE of the unnormalized predictions on the three (not augmented) subsets (i.e., training, validation, test) are evaluated to accurately asses the model's fitness.

\paragraph{Production run}
For the final production run, first, the validation set is merged with the test set (i.e., $n_\mathrm{data}^\mathrm{val}=40$). Second, an ensemble of ANNs is trained with the final set of HPs (see Section~\ref{ssec:hpStudy}) to address the data splitting induced variance (potential bias) in the model's training phase \cite{vanpoucke_small_2020}. The mean values of the ANN ensemble predictions are finally output. Third, the machine learning model is used to predict the case study described in Section~\ref{ssec:setup}. The considered input Ar$^+$ ion energy range (i.e., 3-300 eV) and incident Al neutral to Ar$^+$ ion flux ratios (i.e., 0 to 1) are sampled more continuously (i.e., 1000 steps along each axis). The predicted output surface state $S_\mathrm{s}^\prime$ after the 1st dose is fed as input to the model for the 2nd dose of impinging particles. 

\subsection{Physics-separating artificial neural network}
\label{ssec:model}

In the following, the physical phenomena or processes to be modeled are subdivided into smaller physically disjoint contributions, each of which described by their own individual ANN model. Once established, the combination of submodels form a physics-separating neural network (PSNN) for the intended application. One advantage of such a subdivision is that it allows for the combination of dependent but possibly unavailable sets of descriptors. The model and its training can be set up to depend solely on available descriptors (e.g., measurable quantities), whereas unavailable descriptors (e.g., non measurable or dependent quantities based on theory) are hidden as internal system states. The PSNN applied in this work follows this idea. It consists of two connected regression machine learning models using conditional variational autoencoders (CVAEs) architectures.

\begin{figure}
\includegraphics[width=8cm]{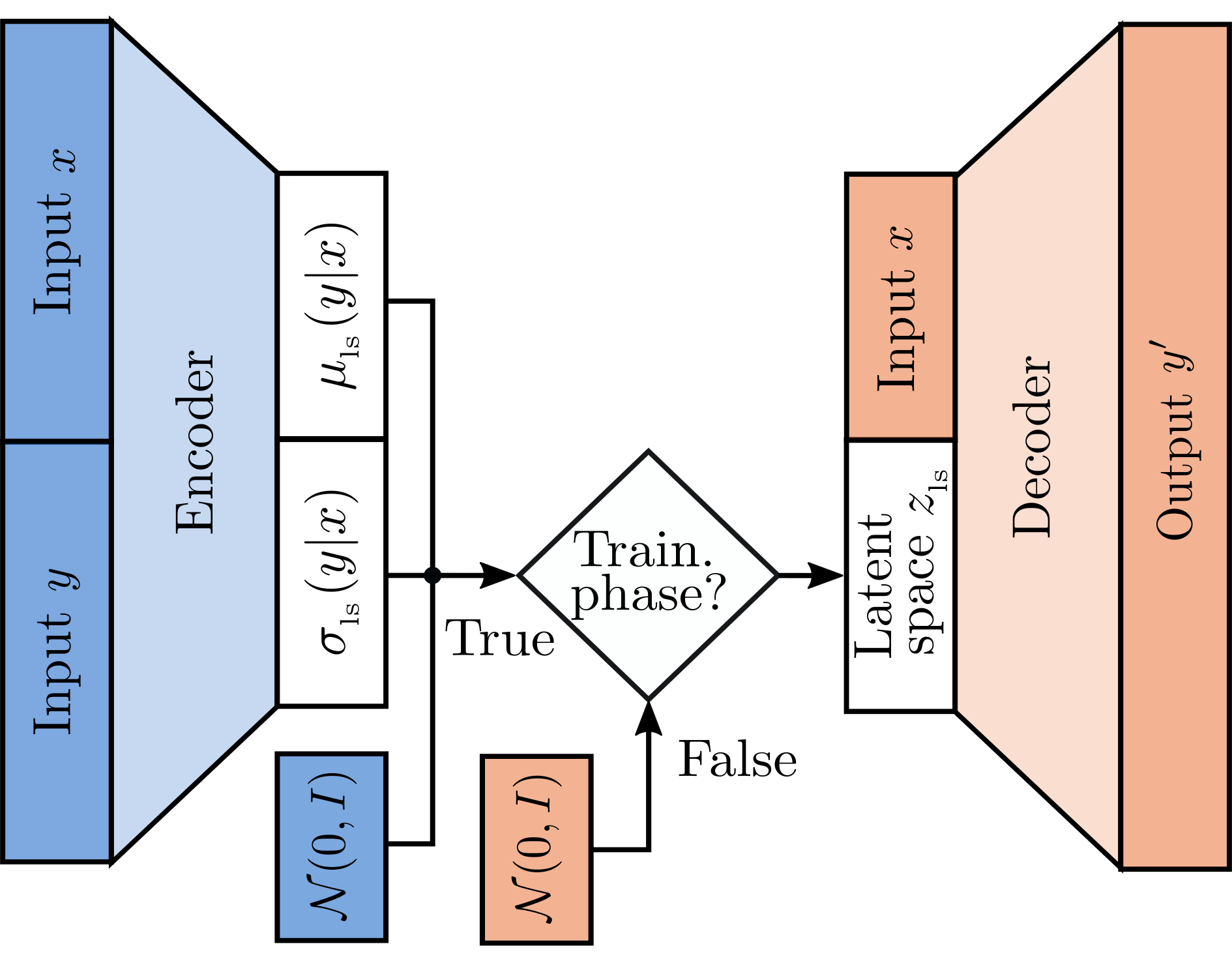}
\caption{Schematic of the CVAE structure. Input variables enter from the left and predictions are extracted on the right of the graph. The coordinates $z_\mathrm{ls}$ of the latent space are indicated by the center white box.}
\label{fig:CVAE_Schematic}
\end{figure}

A CVAE maps an input $x$ to an output $y$ \cite{sohn_learning_2015,doersch_tutorial_2021}. The information flow is depicted schematically in Figure~\ref{fig:CVAE_Schematic}. Model predictions are marked by primes (e.g., $y^\prime$).

First, the training phase is described (blue and orange elements in Figure~\ref{fig:CVAE_Schematic}). An encoder (blue) that is conditioned on $x$ compresses $y|x$ to a $n_\mathrm{ls}$ dimensional latent space (ls) with coordinates $z_{\mathrm{ls}}$. The encoder parameterizes the distribution of training samples by standard deviations $\sigma_{\mathrm{ls}}$ and mean values $\mu_{\mathrm{ls}}$ \cite{sohn_learning_2015,doersch_tutorial_2021}. (Pseudo-)random samples following corresponding normal distributions $\mathcal{N}(\mu_{\mathrm{ls}}, \sigma_{\mathrm{ls}})$ are generated by applying the reparameterization trick \cite{kingma_auto-encoding_2013,higgins_beta-vae_2016} to allow a deterministic backpropagation of error gradients.
%However, this sampling is factually implemented external to the ANN by considering (pseudo-)random samples from a standard normal distribution $\mathcal{N}(0,1)$ as an input. These samples are scaled with the standard deviations $\sigma_{\mathrm{ls}}$ and shifted by the mean values $\mu_{\mathrm{ls}}$. This allows for training with deterministic backpropagation of error gradients and is referred to as reparameterization trick \cite{kingma_auto-encoding_2013,higgins_beta-vae_2016}.
The obtained realizations resemble points in the latent coordinate space $z_{\mathrm{ls}}$. This latent space functions as a bottleneck for the information flow and enhances the model's generalization capabilities. The sampled latent space variables $z_{\mathrm{ls}}|y,x$ project the output from the encoder (blue) to the input of the decoder (orange). The decoder is further conditioned on the input $x$. It finally transforms the inputs $z_{\mathrm{ls}}|y,x$ and $x$ and predicts a representation of the output $y^\prime$, trying to reconstruct $y$. This reconstruction is governed by a reconstruction loss, e.g., given by the mean absolute error (MAE) $\left|y-y^\prime\right|$. During the training phase, the network initially resembles an ordinary $\beta$-variational autoencoder ($\beta$-VAE) whose encoder and decoder are conditioned on the input $x$.

The data transformation through multivariate normal distributions structures the information from the ANN inputs ($x$ and $y$) and distributes it over the latent space coordinates $z_{\mathrm{ls}}$. Initially, conditioned projections $z_{\mathrm{ls}}|y,x$ are distributed with respective standard deviations and mean values. They are subsequently forced toward a common standard normal distribution due to the consideration of an additional Kullback-Leibler (KL) divergence, which pushes $\mathcal{N}(\mu_{\mathrm{ls}}, \sigma_{\mathrm{ls}})$ toward $\mathcal{N}(0,1)$. To achieve this, the KL loss is multiplied with a HP $\beta$ \cite{kingma_auto-encoding_2013, rezende_stochastic_2014}. Simulated annealing is applied to this loss contribution, gradually scaling it with a factor of $10^{-3}$ to 1 over the course of ten epochs (i.e., per batch over a log-axis).

Notably, the encoder (blue elements in Figure~\ref{fig:CVAE_Schematic}) is dismissed post training for prediction (e.g., on the validation or test set). The latent space variables $z_{\mathrm{ls}}$ are then sampled directly from a standard normal distribution. In the end, the decoder (orange elements in Figure~\ref{fig:CVAE_Schematic}) represents the final regression model \cite{sohn_learning_2015,doersch_tutorial_2021}.

\begin{figure}
\includegraphics[width=8cm]{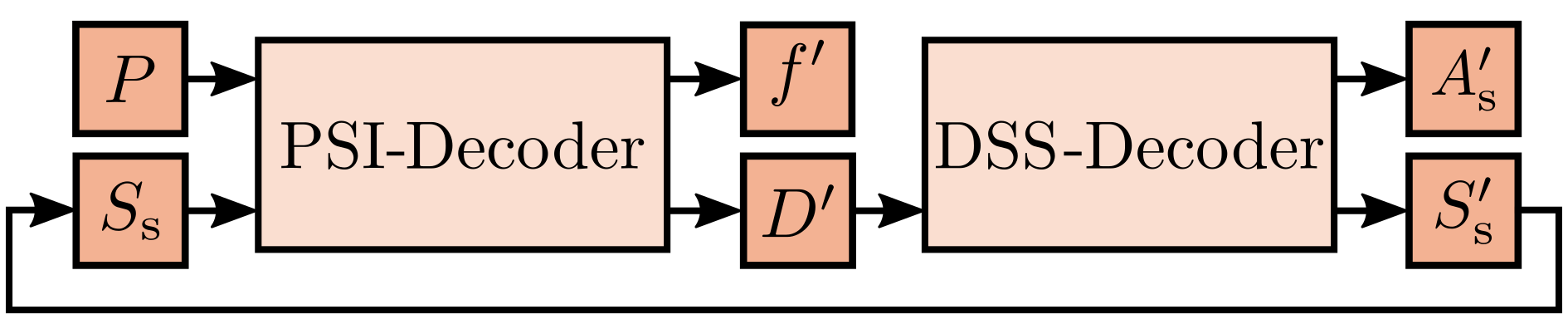}
\caption{Schematic of the PSNN structure and information flow per dose of impinging particles. Input variables enter from the left and predictions are extracted on the right of the graph.}
\label{fig:Stacked_CVAE}
\end{figure}

\paragraph{PSNN training and inference}
The PSNN combines two CVAE decoders for inference as depicted schematically in Figure~\ref{fig:Stacked_CVAE}. PSI (i.e., ion bombardment induced sputtering, damage formation) and damage to surface state conversions (DSS; i.e., defect structure related changes of experimentally attainable surface properties) are modeled by the PSI decoder and the DSS decoder, respectively. The reasoning to separate the surface interaction physics in two ANNs is that the defect structure $D$ is not experimentally accessible and, hence, may be hidden as an internal state in the PSNN. This decision is built on the hypothesis that $D$ as well as $S_\mathrm{s}$ are equivalently sufficient to describe the system state appropriately. Finally, the DSS decoder output $S_\mathrm{s}^\prime$ may be fed to the PSI decoder as input for the next dose of impinging particles, establishing a recurrent connection and assuming that $S_\mathrm{s}$ describes the system state sufficiently.

Notably, while prediction proceeds utilizing only the CVAE decoder networks, training is conducted with the complete CVAE networks. For the training it is moreover fundamental to combine both CVAEs to a single PSNN. Both CVAEs are then trained simultaneously, whereas the encoder networks are provided with the inputs and outputs as shown in Figure~\ref{fig:CVAE_Schematic}. The MAEs of each output ($f^\prime$, $D^\prime$ for PSI and $A_\mathrm{s}^\prime$, $S_\mathrm{s}^\prime$ for DSS) are backpropagated through the PSNN all the way to its inputs. If reference values for the internal descriptors (i.e., $D$) would not be available, e.g., for hypothetical experimental data, the MAE for these descriptors may be set to zero, suspending error backpropagation.

\paragraph{PSI-CVAE}
Both sets of descriptors $P$ and $S_\mathrm{s}$ for the regression task as well as $f$ and $D$ for the autoencoding task are provided to the PSI-CVAE as input. It predicts the ion bombardment induced defect structure $D^\prime$ as well as the EDFs of reflected and sputtered particles $f^\prime$. In its encoder, the EDFs $f$ are passed through three convolutional layers (CLs) with kernel sizes of 9, 7, and 5, filter numbers $n_\mathrm{f}$ (HP) and strides 2 to extract relevant features. These are then concatenated with $D$, $P$ and $S_\mathrm{s}$. They altogether are fed to $n_\mathrm{hl}$ (HP) hidden dense layers (DLs) with $n_\mathrm{npl}$ (HP) nodes per layer resembling an MLP. The output layers of the encoder provide the mean values $\mu_{\mathrm{ls}}$ and standard deviation $\sigma_{\mathrm{ls}}$. The activation functions for any hidden and output layer are set as rectified linear unit (ReLU) and linear, respectively. The CVAEs are chosen to be symmetric and, hence, the decoder is built in the inverse order. CLs are replaced with upsampling transposed convolutional layers. A last custom layer enforces the output $y$ of each CVAE to lie between lower and upper bounds $y_\mathrm{lo}$ and $y_\mathrm{hi}$, respectively. If the input to this custom layer is considered as $x$, the constraint is enforced in the two consecutive computations: i) $y_\mathrm{tmp}=x-2(x-y_\mathrm{lo})(x<y_\mathrm{lo})$, ii) $y=y_\mathrm{tmp}-2(y_\mathrm{tmp}-y_\mathrm{hi})(y_\mathrm{tmp}>y_\mathrm{hi})$. The lower bounds $y_\mathrm{lo}$ for all output properties but the biaxial stress $\frac{\sigma_{xx}+\sigma_{yy}}{2}$ and shear stress $\tau_{xy}$ are zero. The upper bounds $y_\mathrm{hi}$ for the Ar concentration $x_\mathrm{Ar}$, the Ar gas porosity $\phi_\mathrm{Ar}$, and the ring statistical properties $P_{\mathrm{N},r}$, $P_{\mathrm{min},r}$, $P_{\mathrm{max},r}\}$ are 100 (relative percentage) and 1 (relative values), respectively. All remaining upper bounds are chosen to approach infinity. A detailed figure of the network structure can be found in the Appendix. 

\paragraph{DSS-CVAE}
The DSS-CVAE is trained to predict the surface properties $A_\mathrm{s}^\prime$ and surface state $S_\mathrm{s}^\prime$ after the ion bombardment on the basis of the provided (or previously predicted) defect structure $D$. The structure of the PSI-CVAE is adopted for the DSS-CVAE, but no convolutional layers are utilized. In the encoder, $D^\prime$, $A_\mathrm{s}$, and $S_\mathrm{s}$ are concatenated directly. The decoder mirrors the encoder network with the addition of the previously mentioned custom layer to enforce output bounds. For training, $D$ is provided as input for the regression task whereas $A_\mathrm{s}$ and $S_\mathrm{s}$ are provided as input for the autoencoding task. A detailed figure of the DSS-CVAE network structure can be found in the Appendix. Notably, each CVAE is assigned with its own unique set of HPs.

\subsection{Hyperparameter study}
\label{ssec:hpStudy}

The 11 HPs to be optimized are listed in Table~\ref{table: HP_params} ($n_\mathrm{HP}=11$). This is achieved by applying an anisotropic self-adaptive evolution strategy with intermediate recombination $(\mu/\mu_\mathrm{I},\lambda)$-$\sigma$SA-ES. HPs are represented by individual genes $g$ and step sizes (mutation strengths) $\sigma$ that altogether form a list that is referred to as chromosome. The effect of the step sizes (mutation strengths) are detailed later. Each individual in the population has got its own chromosome. The overall population size is specified by $\lambda$. The value of each gene $g$ is initialized by sampling randomly in between $g_{\mathrm{lo}}$ and $g_{\mathrm{hi}}$ (integer types are enforced when necessary). The initialization range intervals for all genes are provided in Table~\ref{table: HP_params}. ANNs corresponding to the individual chromosomes are set up, trained, and eventually evaluated on each (not augmented) subset (training set, validation set, test set). A 10-fold MCCV (and finally a 100-fold MCCV) are utilized as outlined in Section~\ref{ssec:data}. The validation loss is used to sort the individuals by their fitness. The best $\mu$ parents are selected and the centroid of their HPs is computed and assigned to the offspring population. This is referred to as intermediate recombination. The genes are then mutated. First, the step sizes (mutation strengths) for gene $\sigma$ are updated following $\sigma_h\rightarrow \sigma \exp\left[\tau^\prime \mathcal{U}(-1,1) + \tau \mathcal{U}(-1,1)\right]$. $\tau^\prime \mathcal{U}(-1,1)$ and $\tau \mathcal{U}(-1,1))$ enable changes on the global and per gene scale, respectively. The uniform distribution $\mathcal{U}(-1,1)$ with a sampling interval of [-1,1) is used instead of a standard normal distribution. The factors $\tau^\prime$ and $\tau$ are defined by $\tau^\prime=\sqrt{2\sqrt{n_\mathrm{HP}}}^{-1}$ and $\tau=\sqrt{2n_\mathrm{HP}}^{-1}$, respectively \citep{back_basic_1994,schwefel_numerische_1977,rozenberg_handbook_2012}. Second, each gene $g$ is updated following $g\rightarrow\sigma\mathcal{U}(-1,1)(g_{\mathrm{hi}}-g_{\mathrm{lo}})/2+g_h$. Necessary constraints are enforced (e.g., integer numbers, $n_\mathrm{hl}\geq1$).

\begin{table}
\caption{The HPs to be optimized, their initialization range, and final values for the PSI-CVAE as well as DSS-CVAE.}
\label{table: HP_params}
\begin{center}
\begin{tabular}{ l   c   c   c}
\hline
HP & Init. range & PSI-CVAE & DSS-CVAE \\
\hline
$\alpha$		& [0.1,1.0] & 4.96 & 4.96 \\
$r_\mathrm{c}$ 	& [1.0,5.0] & 3.40 & 3.40 \\
$n_\mathrm{f}$	& [2,20] & 14 & --\\
$n_\mathrm{hl}$ & [1,5] & 4 & 22 \\
$n_\mathrm{npl}$& [8,128] & 177 & 158 \\
$n_\mathrm{ls}$	& [1,5] & 14 & 3 \\
$\beta$			& [1.0,10.0] &  2.06 & 3.17$\cdot10^4$ \\
\hline
\end{tabular}
\end{center}
\end{table} 

The HP study is initiated with 70 individuals ($\lambda=70$) and reduced by one per generation until eventually a population size of seven is kept constant. The number of parents is determined by $\mu=\lambda/7$ (integer values are enforced) \cite{schwefel1987collective}. Hence, a $(7/7_\mathrm{I},70)$-$\sigma$SA-ES is gradually turned into a $(1,7)$-$\sigma$SA-ES over the course of 63 generations. The $(1,7)$-$\sigma$SA-ES is then continued for the remaining generations.

%%%%%%%%%%%%%%%%%%%%%%%%%%%%%%%%%%%%%%%%%%%%%%%%%%%

\section{Results}
\label{sec:results}

First, the outcome of the HP study and the machine learning model's generalization capability is assessed. Second, the results of the production run with the final set of HPs are discussed concerning the conservation of physical principles observed in the underlying case study \cite{gergs_molecular_2022}. This study focuses on only the most relevant physical properties.

\subsection{Hyperparameter study}
\label{ssec:hpStudy_results}

\begin{figure}
\includegraphics[width=8cm]{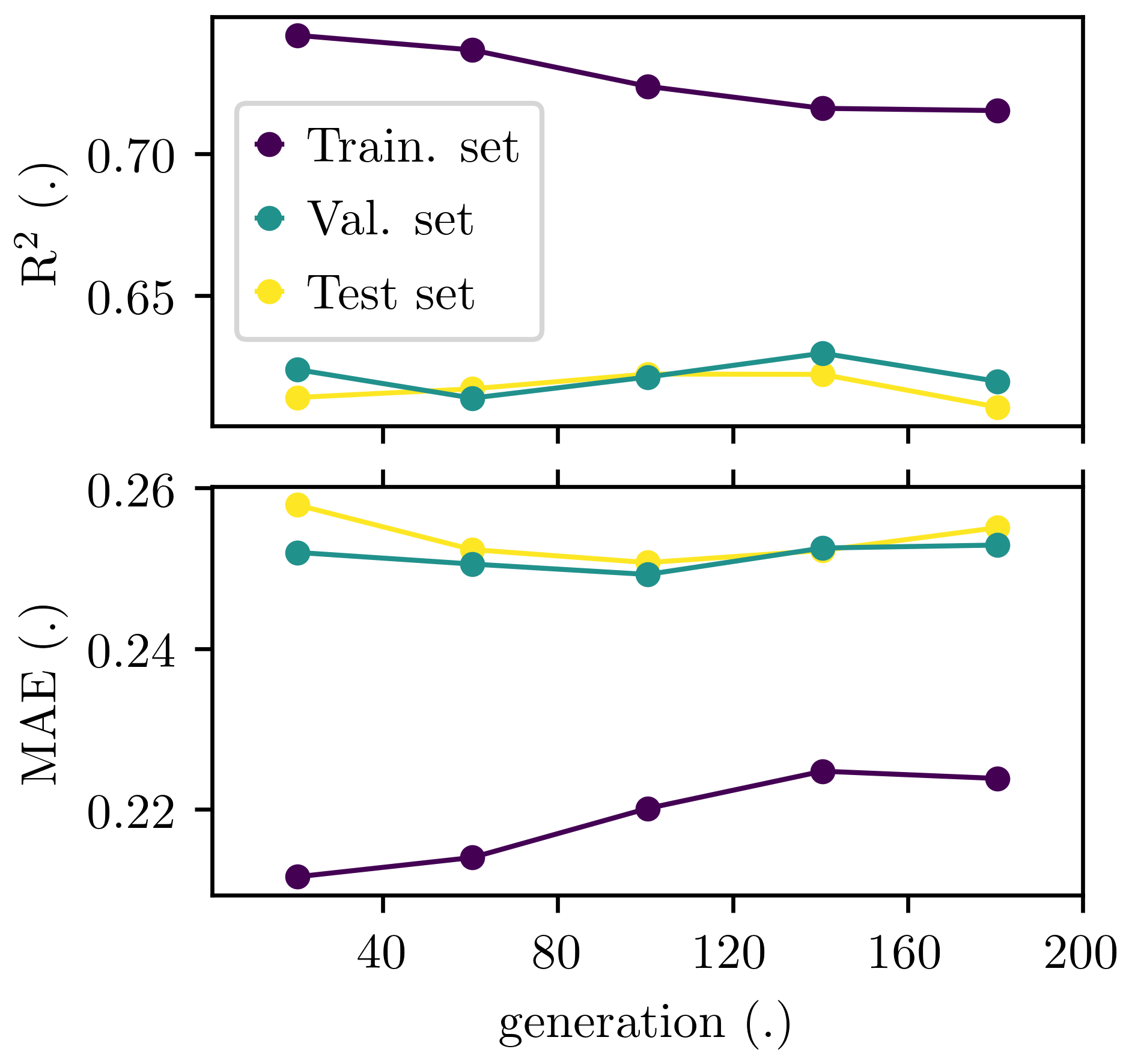}
\caption{Evolution of the MAE and $R^2$ with normalized data and predictions for all three subsets. The metrics are averaged over 40 generations.}
\label{fig:MAE_R2}
\end{figure}

\begin{figure*}
\includegraphics[width=16cm]{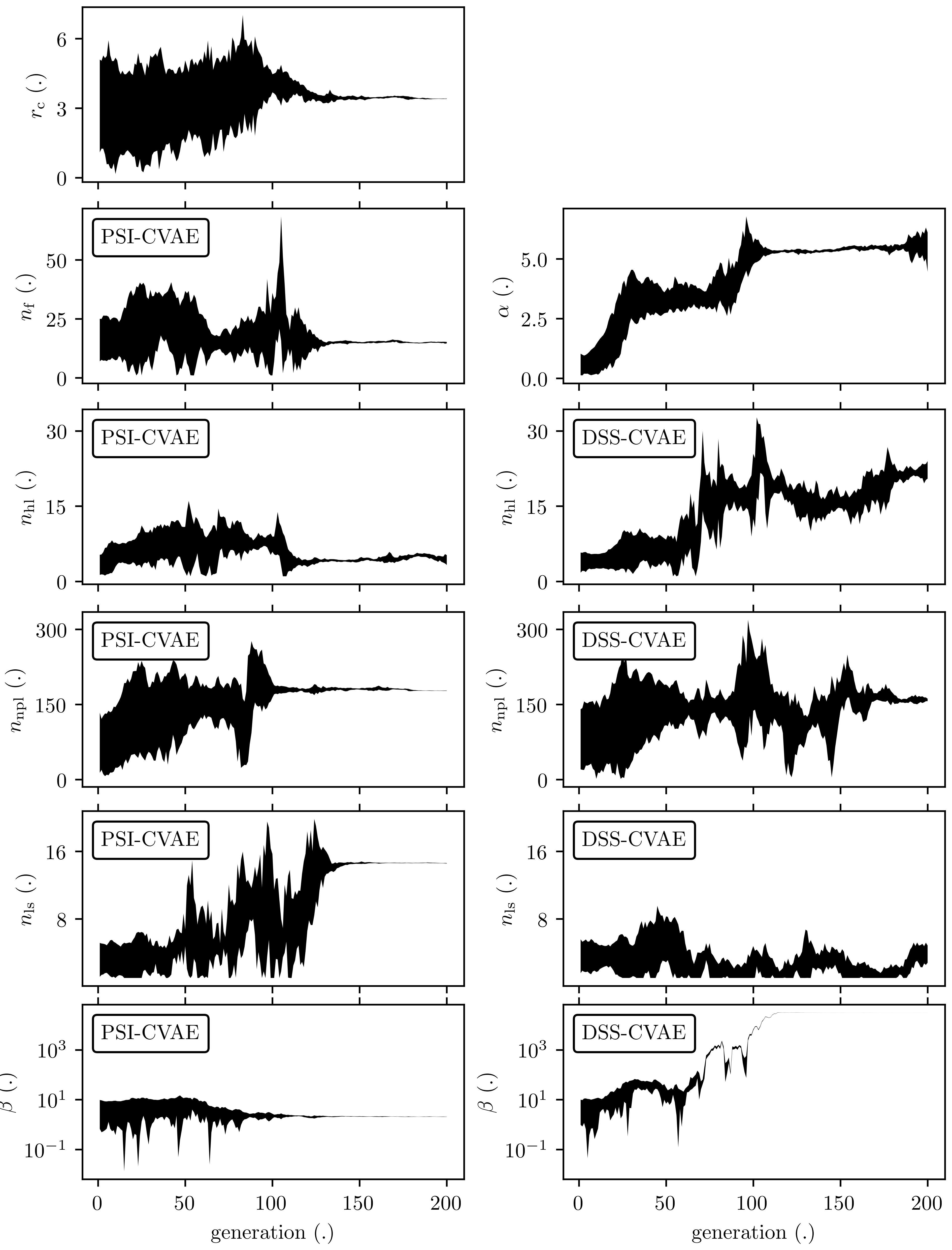}
\caption{The evolution of the HPs' centroids plus/minus the step sizes (mutation strengths) $\sigma_h$ are shown to indicate the search range.}
\label{fig:HPs}
\end{figure*}

For the evolution strategy of HP optimization, the MAE and $R^2$ evolved for 200 and averaged over the course of 40 generations are shown in Figure~\ref{fig:MAE_R2}. The initial deviations between the errors on the validation as well as the test set are removed after 60 generations. The gap to the errors on the training set is reduced within 140 generations. Both findings indicate an improved generalization. Afterwards, the evolution strategy converges. The final values of all HPs can be found in Table~\ref{table: HP_result}, whereas the evolution is depicted in Figure~\ref{fig:HPs}. 

Mixup augmentation provides a hypothesis of linear superposition to the PSNN. An $\alpha=4.96$ value greater one makes data augmentation by means of averaging training set sample pairs ($\lambda$=0.5) the most probable event. $r_\mathrm{c}=3.4$ constrains this augmentation to include only combination of sample pairs up to a radius of 3.4 nearest indices. Hence, the hypothesis outlined before is accepted for this selection. The DSS-CVAE is constructed with a significantly higher number of hidden layers ($n_\mathrm{hl}=$22) than the PSI-CVAE ($n_\mathrm{hl}=$4). While such a complexity is not required to perform the regression operation in an acceptable manner, the high number of degrees of freedom (i.e., weights) facilitates a more robust training and smooth prediction at the cost of a less efficient training (i.e., increased computational cost) \cite{bubeck_universal_2021}. The difference of the PSI-CVAE and DSS-CVAE latent space dimensions (i.e., 14 and 3, respectively) is assumed to be correlated with the vector sizes (i.e., 92 and 23, respectively) input to the encoding hidden layers. The smaller $\beta$ value used for the PSI-CVAE of $\beta=2.06$ may signify the relevance of the autoencoding part of the CVAE, preconditioning the weights (cf. DSS-CVAE $\beta=3.17\cdot 10^4$). 

\begin{table}
\caption{MAE of the unnormalized predictions on all considered properties and data sets. For the EDFs $f_{\mathrm{Al}}$, $f_{\mathrm{Ar}}$ and the connectivity profile $R_{\mathrm{C},r}$, $P_{\mathrm{N},r}$, $P_{\mathrm{min},r}$, $P_{\mathrm{max},r}$, the MAE averaged over the energies and the ring sizes is presented, respectively.}
\label{table: HP_result}
\begin{center}
\begin{tabular}{ l   c   c c}
\hline
Property & Train. set & Val. set & Test set \\
\hline
$f_{\mathrm{Al}}$ ($10^{-3}$)					& 1.31 & 1.54 & 1.56 \\
$f_{\mathrm{Ar}}$ ($10^{-3}$)					& 2.37 & 2.69 & 2.78 \\
$R_{\mathrm{C},r}$ ($10^{-2}$)			& 4.77 & 5.46 & 5.59 \\
$P_{\mathrm{N},r}$ ($10^{-2}$)			& 1.11 & 1.23 & 1.26 \\
$P_{\mathrm{min},r}$ ($10^{-2}$)		& 3.95 & 4.63 & 4.25 \\
$P_{\mathrm{max},r}$ ($10^{-5}$)		& 3.08 & 3.02 & 3.37 \\
$\phi_\mathrm{Ar}$ (\%)						& 0.380 & 0.436 & 0.503 \\
$N_\mathrm{v_{Al}}$ ($10^{19}$ cm$^{-3}$)	& 3.14 & 3.25 & 3.56 \\
$R_\mathrm{RMS}$ (\r A)						& 0.166 & 0.189 & 0.186 \\
$x_\mathrm{Ar}$ (\%)						& 0.239 & 0.283 & 0.313 \\
$\rho$ ($10^{-3}$ g/cm$^3$)			& 4.94 & 5.49 & 5.89 \\
$\frac{\sigma_{xx}+\sigma_{yy}}{2}$ (MPa)	& 52.5 & 55.3 & 58.1 \\
$\tau_{xy}$ (MPa)							& 31.2 & 32.9 & 32.9 \\
\hline
\end{tabular}
\end{center}
\end{table}

The final MAE on the training, the validation, and the test set are presented in Table~\ref{table: HP_result} for all considered properties. The reference values include statistical noise as well as physical fluctuations (i.e., cycle of Ar incorporation, clustering, outgassing) and, hence, do not resemble any kind of ground truth \cite{gergs_molecular_2022}. This provides an inherent lower limit for the MAE. For instance, the shear stress is predicted correctly to equal zero for any incident particle flux by the PSNN. The provided MAE of the shear stress $\tau_{xy}$ in Table~\ref{table: HP_result} describes solely the statistical noise. Hence, the resultant errors on the unknown test set are considered to be sufficiently small when comparing them to the underlying data in terms of magnitude, fluctuation, and noise \cite{gergs_molecular_2022}. Reflection of incident Ar$^+$ ions and outgassing of incorporated Ar atoms lead to an overall higher Ar flux leaving the surface. Thus, a higher MAE is observed for outgoing EDFs of Ar $f_{\mathrm{Ar}}$ than for Al $f_{\mathrm{Al}}$.

\subsection{Production run}
\label{ssec:production_results}

\begin{figure}
\includegraphics[width=8cm]{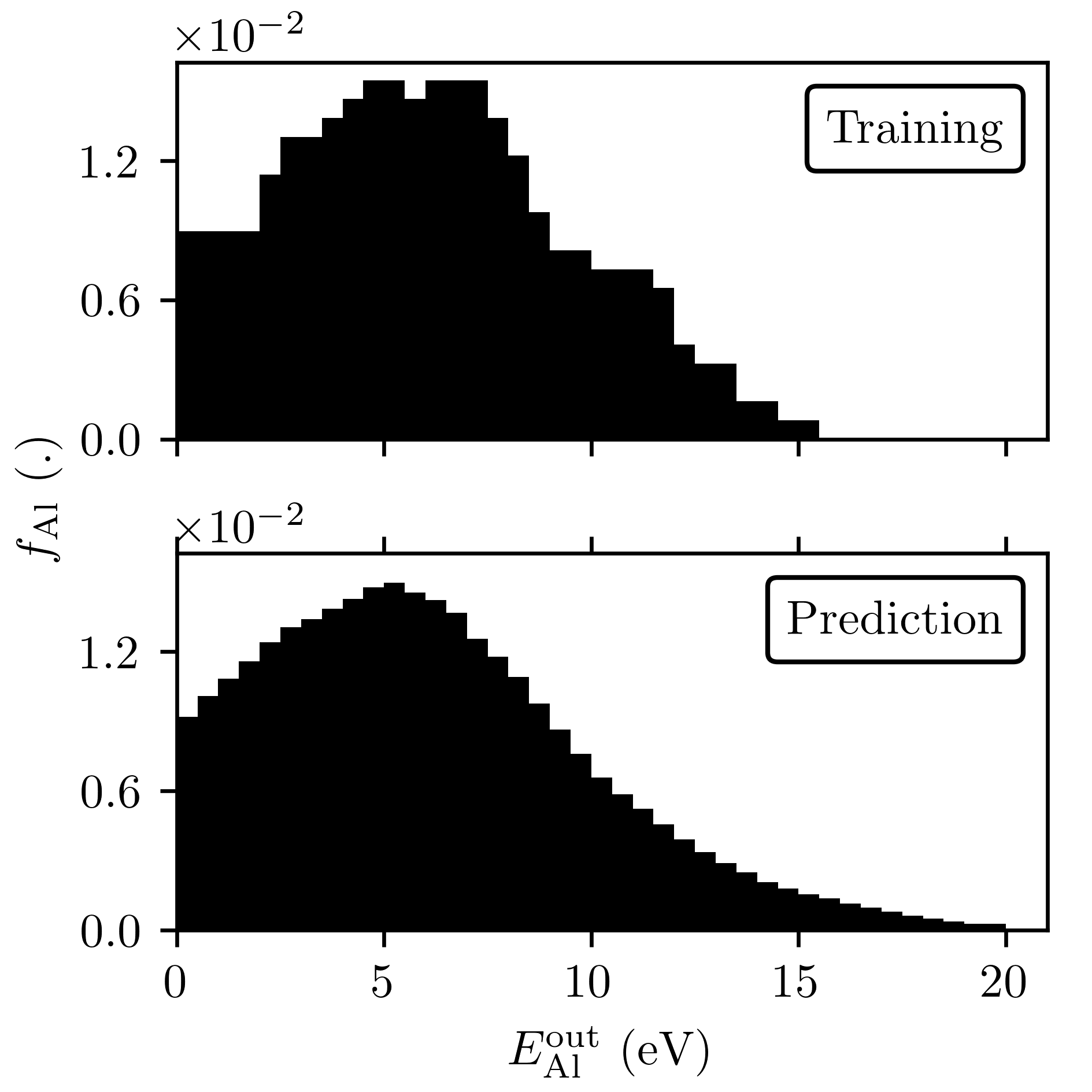}
\caption{The EDF of reflected and sputtered Al atoms. The absolute frequency is normalized by the number of incident Ar$^+$ ions.}
\label{fig:Al_edf}
\end{figure}

The first outputs of the PSNN ensemble are the EDFs $f$ of reflected as well as sputtered Al and Ar atoms. An example for an Al EDF is shown in Figure~\ref{fig:Al_edf} for an incident Ar$^+$ ion energy and Al neutral to Ar$^+$ ion flux ratio of 197 eV and 0.42, respectively. The noise is mitigated and the distribution function matches the physical expectation of a continuous representation. The peak at approximately 5 eV relates to the surface binding energy and is found to be more pronounced than in the displayed reference data.

\begin{figure}
\includegraphics[width=8cm]{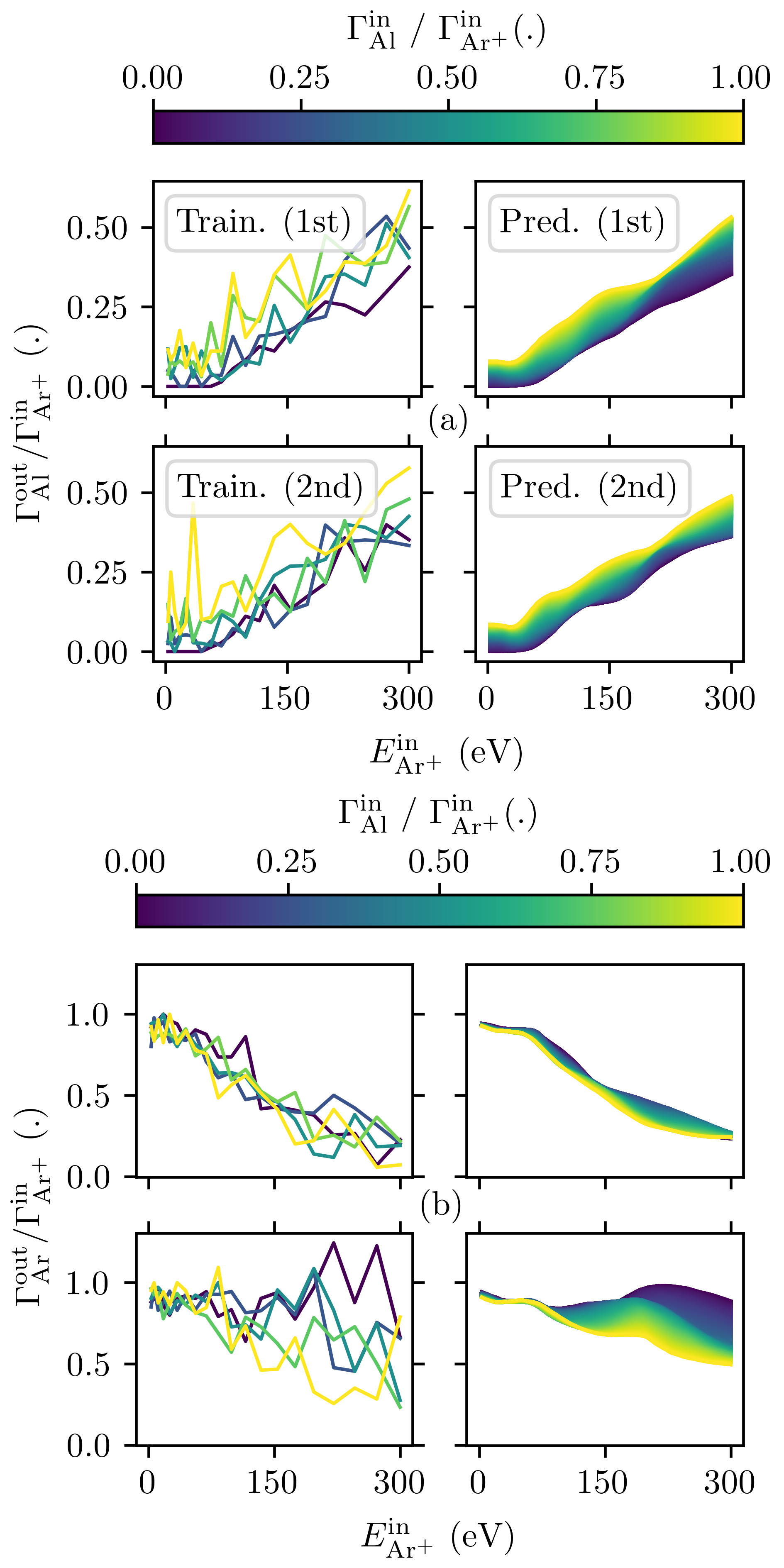}
\caption{(a) Flux ratio $\Gamma_\mathrm{Al}^\mathrm{out}/\Gamma_\mathrm{Ar^+}^\mathrm{in}$ and (b) $\Gamma_\mathrm{Ar}^\mathrm{out}/\Gamma_\mathrm{Ar^+}^\mathrm{in}$ is shown as a function of the Ar$^+$ ion energy $E_\mathrm{Ar^+}^\mathrm{in}$ and incident flux ratio $\Gamma_\mathrm{Al}^\mathrm{in}/\Gamma_\mathrm{Ar^+}^\mathrm{in}$ for the 1st and 2nd dose of impinging particles. The training data and model predictions are compared with each other.}
\label{fig:yields}
\end{figure}

The Al EDFs integrated over the energy axis yield the $\Gamma_\mathrm{Al}^\mathrm{out}/\Gamma_\mathrm{Ar^+}^\mathrm{in}$ flux ratios. Prediction and reference data are presented in Figure~\ref{fig:yields} (a). Fluctuations are reduced significantly but persist weakly. Above a threshold of approximately $E_\mathrm{Ar^+}^\mathrm{in}\approx50$ eV an expected monotonously increasing dependence is observed. The threshold is commonly referred to as sputtering threshold and also relates to the surface binding energy. A raise of the $\Gamma_\mathrm{Al}^\mathrm{out}/\Gamma_\mathrm{Ar^+}^\mathrm{in}$ flux ratio with increasing incident $\Gamma_\mathrm{Al}^\mathrm{in}/\Gamma_\mathrm{Ar^+}^\mathrm{in}$ flux ratios is predicted by the PSNN, independent of the Ar$^+$ ion energy. This is related to the reflection of Al atoms, which in the hybrid RMD/tfMC case study are found to be Ar$^+$ ion energy independent too. The hybrid RMD/tfMC and PSNN simulations predict the mean reflection ratio for incident Al neutrals to equal 9.97 \% and 9.57 \%, respectively (assuming $E_\mathrm{Ar^+}^\mathrm{in}=3$\,eV to rule out sputtering of Al surface atoms and separate the reflection of incident Al atoms). The reflection and sputtering of Al atoms are predicted accurately \cite{gergs_molecular_2022}.

In the case study that provided the reference data for training the machine learning model, an ongoing cycle of Ar incorporation, clustering, and outgassing was revealed \cite{gergs_molecular_2022}. This cycle was found to cause correlated fluctuations for the $\Gamma_\mathrm{Ar}^\mathrm{out}/\Gamma_\mathrm{Ar^+}^\mathrm{in}$ flux ratio and other surface properties (e.g., Ar concentration $x_\mathrm{Ar}$). The reference and predicted $\Gamma_\mathrm{Ar}^\mathrm{out}/\Gamma_\mathrm{Ar^+}^\mathrm{in}$ flux ratio are shown in Figure~\ref{fig:yields} (b). Physical fluctuations and statistical noise are reduced effectively by the PSNN. A good agreement between visible trends in the reference data and predictions is achieved. After the 1st dose of impinging particles, a negligible dependence of $\Gamma_\mathrm{Ar}^\mathrm{out}/\Gamma_\mathrm{Ar^+}^\mathrm{in}$ on the incident $\Gamma_\mathrm{Al}^\mathrm{in}/\Gamma_\mathrm{Ar^+}^\mathrm{in}$ flux ratio is observed. Higher ion energies cause a deeper and more likely Ar implantation. The majority of incorporated Ar atoms may cluster but do not yet outgas. After the 2nd dose, a saturation of the surface slab and outgassing events lead to the predicted increase of $\Gamma_\mathrm{Ar}^\mathrm{out}/\Gamma_\mathrm{Ar^+}^\mathrm{in}$. Its dependence on the incident $\Gamma_\mathrm{Al}^\mathrm{in}/\Gamma_\mathrm{Ar^+}^\mathrm{in}$ flux ratio, which is observed now, is reasoned twofold: i) A more pure Ar$^+$ ion flux (i.e., $\Gamma_\mathrm{Al}^\mathrm{in}/\Gamma_\mathrm{Ar^+}^\mathrm{in}\rightarrow0$) consists of a higher number of impinging Ar$^+$ ions leading to faster saturation of the surface slab; ii) An eased (hindered) outgassing for a capping surface layer that becomes thinner (thicker) over time due to a number of sputtered Al atoms that exceeds (falls below) the number of built-in Al atoms per bombarding Ar$^+$ ion, $\Gamma_\mathrm{Al}^\mathrm{in}/\Gamma_\mathrm{Ar^+}^\mathrm{in}<\Gamma_\mathrm{Al}^\mathrm{out}/\Gamma_\mathrm{Ar^+}^\mathrm{in}$ $\left(\Gamma_\mathrm{Al}^\mathrm{in}/\Gamma_\mathrm{Ar^+}^\mathrm{in}>\Gamma_\mathrm{Al}^\mathrm{out}/\Gamma_\mathrm{Ar^+}^\mathrm{in}\right)$. This may be identified from Figure~\ref{fig:yields} (a) and Figure~\ref{fig:yields} (b) \cite{gergs_molecular_2022}.

\begin{figure*}
\includegraphics[width=16cm]{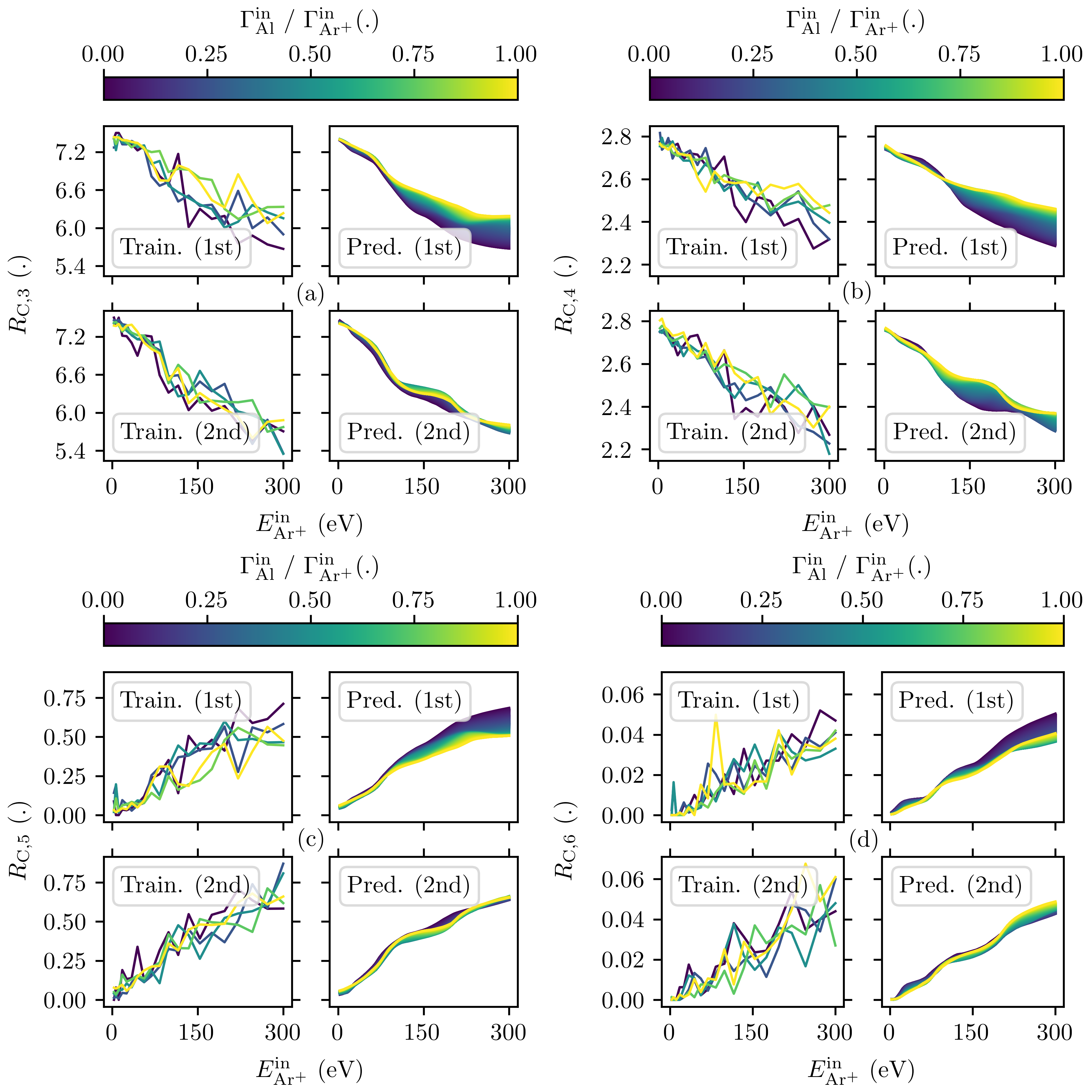}
\caption{(a) $R_\mathrm{C,3}$, (b) $R_\mathrm{C,4}$, (c) $R_\mathrm{C,5}$, and (d) $R_\mathrm{C,6}$ are shown as a function of the Ar$^+$ ion energy $E_\mathrm{Ar^+}^\mathrm{in}$ and incident flux ratio $\Gamma_\mathrm{Al}^\mathrm{in}/\Gamma_\mathrm{Ar^+}^\mathrm{in}$ for the 1st and 2nd dose of impinging particles. The training data and model predictions are compared with each other.}
\label{fig:Rcs}
\end{figure*}

The ion bombardment induced defect structure is quantified by means of the ring statistical connectivity profile (see Section~\ref{ssec:setup}) \cite{drabold2005models,cobb1996ab,zhang2000structural}. Out of which $R_{\mathrm{C},r}$ was found to be the most relevant property and is shown in Figure~\ref{fig:Rcs} for $r\in[3,6]$. Rings with three or four members are sufficient to describe the ideal face-centered-cubic (fcc) Al lattice structure. Rings with five nodes are related to interstitials and surface reconstructions, whereas rings with six members relate to vacancies and Ar atoms occupying vacant Al sites, respectively \cite{gergs_molecular_2022}. Hence, $R_{\mathrm{C},3}$ and $R_{\mathrm{C},4}$ ($R_{\mathrm{C},5}$ and $R_{\mathrm{C},6}$) are found to monotonously decrease (increase) for higher Ar$^+$ ion energies. An effect of the incident flux ratio $\Gamma_\mathrm{Al}^\mathrm{in}/\Gamma_\mathrm{Ar^+}^\mathrm{in}$ is predicted for Ar$^+$ ion energy $E_\mathrm{Ar^+}^\mathrm{in}$ greater than approximately 100 eV. This threshold was observed also in the underlying RMD case study and is due to the minimal kinetic energy required to cause the ongoing cycle of Ar incorporation, clustering, and outgassing \cite{gergs_molecular_2022}. Persistent fluctuations are attributed to the varying states within this cycle after the 1st and 2nd dose of impinging particles. Hence, the herein considered properties are intrinsically correlated, which is thus assumed to be accepted by the PSNN. The remaining properties of the connectivity profile (i.e., $P_{\mathrm{N},r}$, $P_{\mathrm{min},r}$, $P_{\mathrm{max},r}$ \cite{drabold2005models,cobb1996ab,zhang2000structural}) are in excellent agreement with the reference data (see Table~\ref{table: HP_result} and Section~\ref{ssec:hpStudy_results}).

\begin{figure}
\includegraphics[width=8cm]{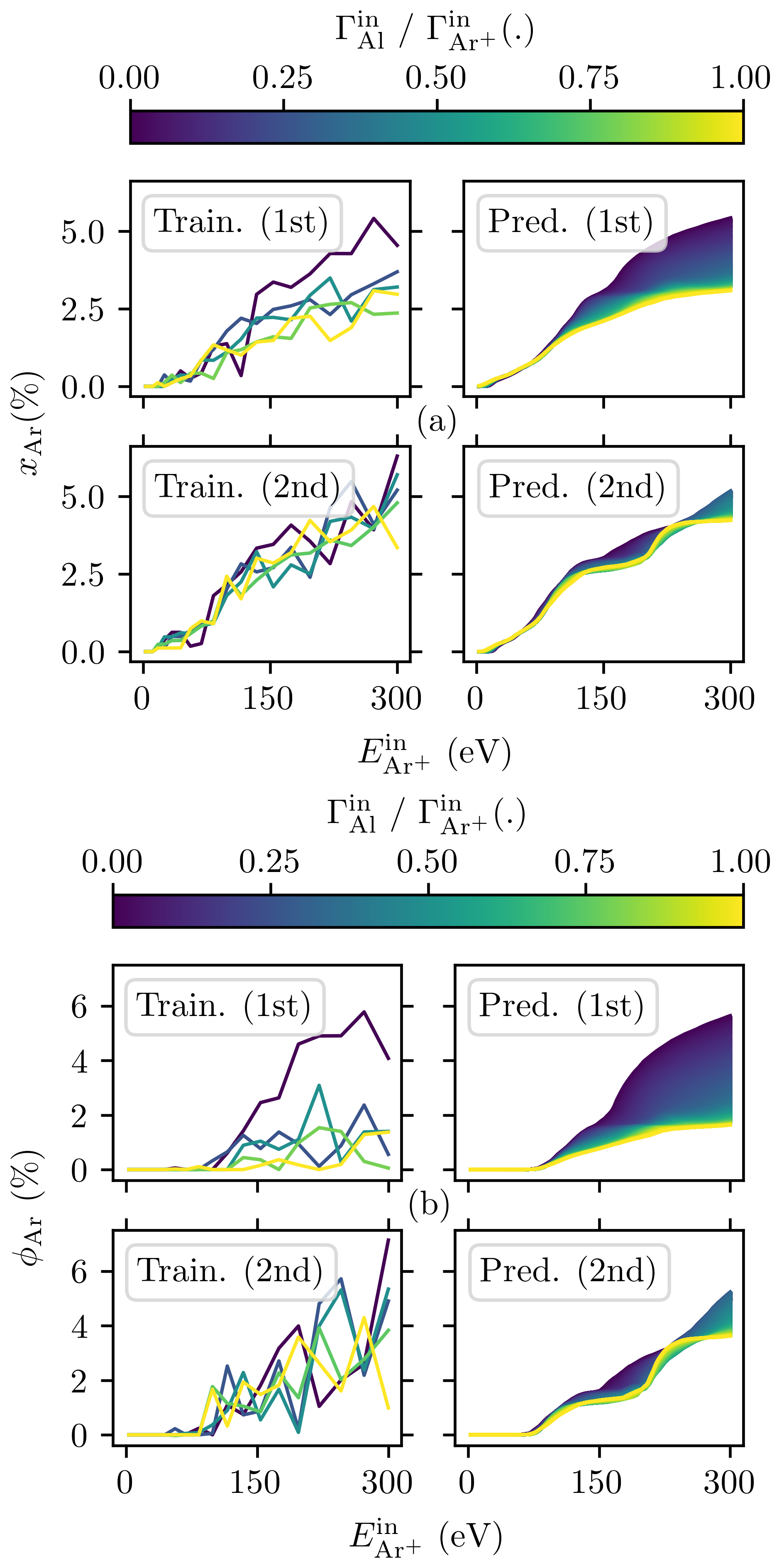}
\caption{(a) Ar concentration $x_\mathrm{Ar}$ and (b) Ar gas porosity $\phi_\mathrm{Ar}$ are shown as a function of the Ar$^+$ ion energy $E_\mathrm{Ar^+}^\mathrm{in}$ and flux ratio $\Gamma_\mathrm{Al}^\mathrm{in}/\Gamma_\mathrm{Ar^+}^\mathrm{in}$ for the 1st and 2nd dose of impinging particles. The training data and model predictions are compared with each other.}
\label{fig:Ar_stoi_gasP}
\end{figure}

The predicted defect structure is used to further predict the corresponding surface state. The mass density (not shown) and Ar concentration $x_\mathrm{Ar}$ (shown in Figure~\ref{fig:Ar_stoi_gasP} (a)) entail the same threshold at approximately 100 eV for the Ar cycle. Their dependence on the incident $\Gamma_\mathrm{Al}^\mathrm{in}/\Gamma_\mathrm{Ar^+}^\mathrm{in}$ flux ratio for ion energies greater than approximately 100 eV is similarly observed. A more pure Ar$^+$ ion flux (i.e., $\Gamma_\mathrm{Al}^\mathrm{in}/\Gamma_\mathrm{Ar^+}^\mathrm{in} \rightarrow 0$) leads to a higher number of impinging Ar$^+$ ions, saturating the surface slab more effectively \cite{gergs_molecular_2022}. The predicted biaxial and shear stresses do not hold such a dependency on the incident $\Gamma_\mathrm{Al}^\mathrm{in}/\Gamma_\mathrm{Ar^+}^\mathrm{in}$ flux ratio for any herein considered Ar$^+$ ion energy (consistent with \cite{gergs_molecular_2022}). Notably, fluctuations are mitigated for all predicted surface state properties (i.e., $\rho^\prime$, $x_\mathrm{Ar}^\prime$, $\frac{\sigma_{xx}^\prime+\sigma_{yy}^\prime}{2}$, $\tau_{xy}^\prime$) which after the 1st dose are fed as input to the PSNN for the 2nd dose of impinging particles.

The machine learning model is found to reliably describe the Ar concentration $x_\mathrm{Ar}$. The Ar clustering is assessed next by comparing the reference with the predicted Ar gas porosity $\phi_\mathrm{Ar}$ in Figure~\ref{fig:Ar_stoi_gasP} (b). The comparison of the ML model predicted $x_\mathrm{Ar}^\prime$ and $\phi_\mathrm{Ar}^\prime$ signify the correlation of the Ar uptake by and spatial distribution within the surface slab. The expected threshold at approximately 100 eV and the incident Al neutral to Ar$^+$ ion flux ratio dependence for higher ion energies is accurately reflected by the PSNN. The 2nd dose leads to an increased volumetric partitioning due to the additional Ar$^+$ ions bombarding the surface, forming larger clusters \cite{gergs_molecular_2022}. The predictions' agreement with the reference data may be assessed more quantitatively by considering Table~\ref{table: HP_result} in Section~\ref{ssec:hpStudy_results}.

%%%%%%%%%%%%%%%%%%%%%%%%%%%%%%%%%%%%%%%%%%%%%%%%%%%
\section{Conclusion}
\label{sec:conclusion}

This work aims further advance the development of a data-driven plasma surface interaction model \cite{gergs_efficient_2022}. It is proposed that a common description of the target and the substrate plasma-surface interactions may be attained, which builds on data obtained from hybrid RMD/tfMC simulations. This description entails a comprehensive system state (i.e., surface state $S_\mathrm{s}$, defect structure $D$). It is established through a physics-separating artificial neural network that conceptually allows to consistently combine mixed input data with different descriptive power. In the future, a combination of simulation data (e.g., $S_\mathrm{s}$, $D$) with experimental data (e.g., only $S_\mathrm{s}$) may be adopted. The PSNN is optimized by applying an anisotropic self-adaptive evolution strategy with intermediate recombination.

The PSNN is found to accurately predict the surface state $S_\mathrm{s}^\prime$ and its properties $A_\mathrm{s}^\prime$ taking only the ion bombardment induced defect structure $D^\prime$ into account. Moreover, it describes accurately the first stages of a transient surface state evolution (i.e., $S_\mathrm{s}$ $\rightarrow$ $S_\mathrm{s}^\prime$ after the 1st dose $\rightarrow$ $S_\mathrm{s}^\prime$ after the 2nd dose). This indicates altogether that the assumption (hypothesis) of a complete system state description by either the defect structure $D$ or surface state $S_\mathrm{s}$ to be justified \cite{karimi_aghda_unravelling_2021}.

The proposed machine learning model and procedure effectively adopts the physical fluctuations for an ongoing cycle of Ar incorporation, clustering, and outgassing. At the same time it reduces the statistical variations of the atomic system that previously hindered the interpretation of the hybrid RMD/tfMC case study. Observations, physical principles, and conclusions revealed in the referenced case study are found to be preserved by the PSNN \cite{gergs_molecular_2022}. Hence, the effect of the plasma (i.e., $P=\{E_{\mathrm{Ar^+}},\Gamma_{\mathrm{Al}}^\mathrm{in}/\Gamma_{\mathrm{Ar^+}}^\mathrm{in}\}$) on the relevant surface properties (i.e., $D$, $S_\mathrm{s}$, $A_\mathrm{s}$) can be revealed more clearly.
 
The PSNN has learned to generalize on a limited data set consisting of only 200 samples. The proposed constrained mixup augmentation is assumed to address this issue by establishing a more robust basis for training the model. Such an effect has been shown for other methods that virtually (artifically) extend the database \cite{oviedo_fast_2019,li_genetic_2014,choudhury_artificial_2011}. It can be perceived as a hypothesis of linear superposition that is provided to the machine learning model. This hypothesis is accepted in this work for training set sample pairs along the Ar$^+$ ion energy, the Al neutral to Ar$^+$ ion flux ratio, and the dose axis with an indices radius equal less than 3.4. 

In summary, the impingement of up to 15280 particles was simulated to generate a small data set that consists of only 200 samples \cite{gergs_molecular_2022}. The system trajectories are found to be predicted accurately for a few ($\lesssim2$) steps (particle doses). This is enabled by the external recurrent connection and the corresponding feedback of the surface state $S_\mathrm{s}$. The PSNN model representation is assumed to not allow for a continued evolution ($\gg 2$ particle doses). This issue is related to the herein considered four dimensional surface state and the 16 dimensional defect structure (probed just 200 times). The transient system state requires extrapolation, which may only be remedied by increasing the number of impinging particles with immense computational costs. This aspect is even more challenging when considering reactive systems (e.g., the sputter deposition of AlN thin films) due to the eventually required application of variable charge models (e.g., QTE, ACKS2) in the RMD or hybrid RMD/tfMC simulations \cite{gergs_generalized_2021,verstraelen2013acks2,verstraelen2014direct}.

%%%%%%%%%%%%%%%%%%%%%%%%%%%%%%%%%%%%%%%%%%%%%%%%%%%
\section*{Acknowledgement}
Funded by the Deutsche Forschungsgemeinschaft (DFG, German Research Foundation) -- Project-ID 138690629 -- TRR 87 and -- Project-ID 434434223 -- SFB 1461.

%%%%%%%%%%%%%%%%%%%%%%%%%%%%%%%%%%%%%%%%%%%%%%%%%%%
\section*{Data Availability}

The data that support the findings of this study are available from the corresponding author upon reasonable request.

%%%%%%%%%%%%%%%%%%%%%%%%%%%%%%%%%%%%%%%%%%%%%%%%%%%

\section*{ORCID}
\noindent
T. Gergs: \url{https://orcid.org/0000-0001-5041-2941} \\
T. Mussenbrock: \url{https://orcid.org/0000-0001-6445-4990} \\
J. Trieschmann: \url{https://orcid.org/0000-0001-9136-8019}

%%%%%%%%%%%%%%%%%%%%%%%%%%%%%%%%%%%%%%%%%%%%%%%%%%%

\clearpage
\appendix
\section*{Appendix}

\begin{figure*}[h]
\includegraphics[width=16cm]{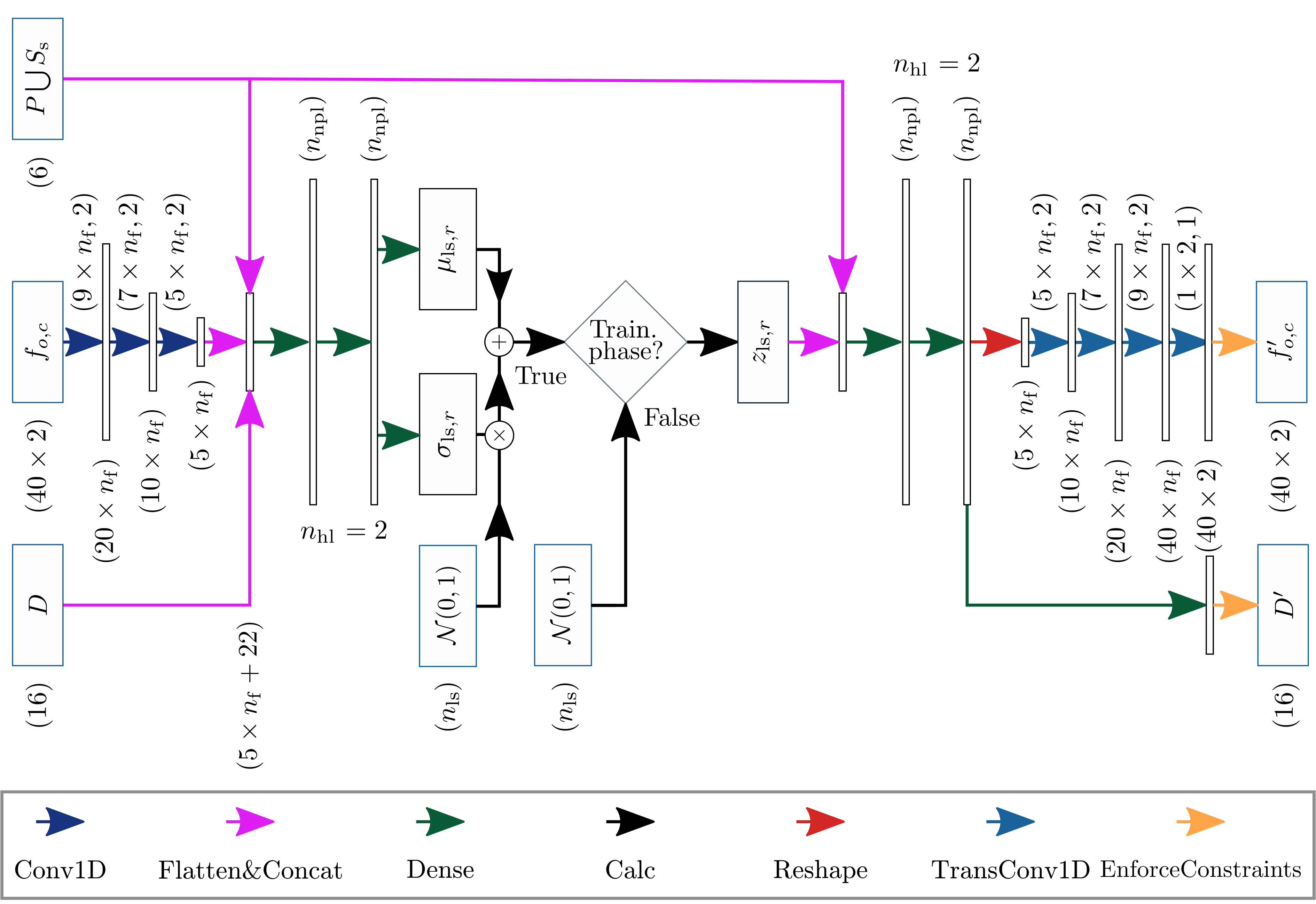}
\caption{Schematic of the PSI-CVAE network structure. The shape of the data is provided in parenthesis. Machine learning operations are indicated by colored arrows. Specifications of the convolutional operations are presented atop the arrows (i.e., kernel size $\times$ number of filters $n_\mathrm{f}$, stride).}
\label{fig:CVAE1}
\end{figure*}

\begin{figure*}
\includegraphics[width=16cm]{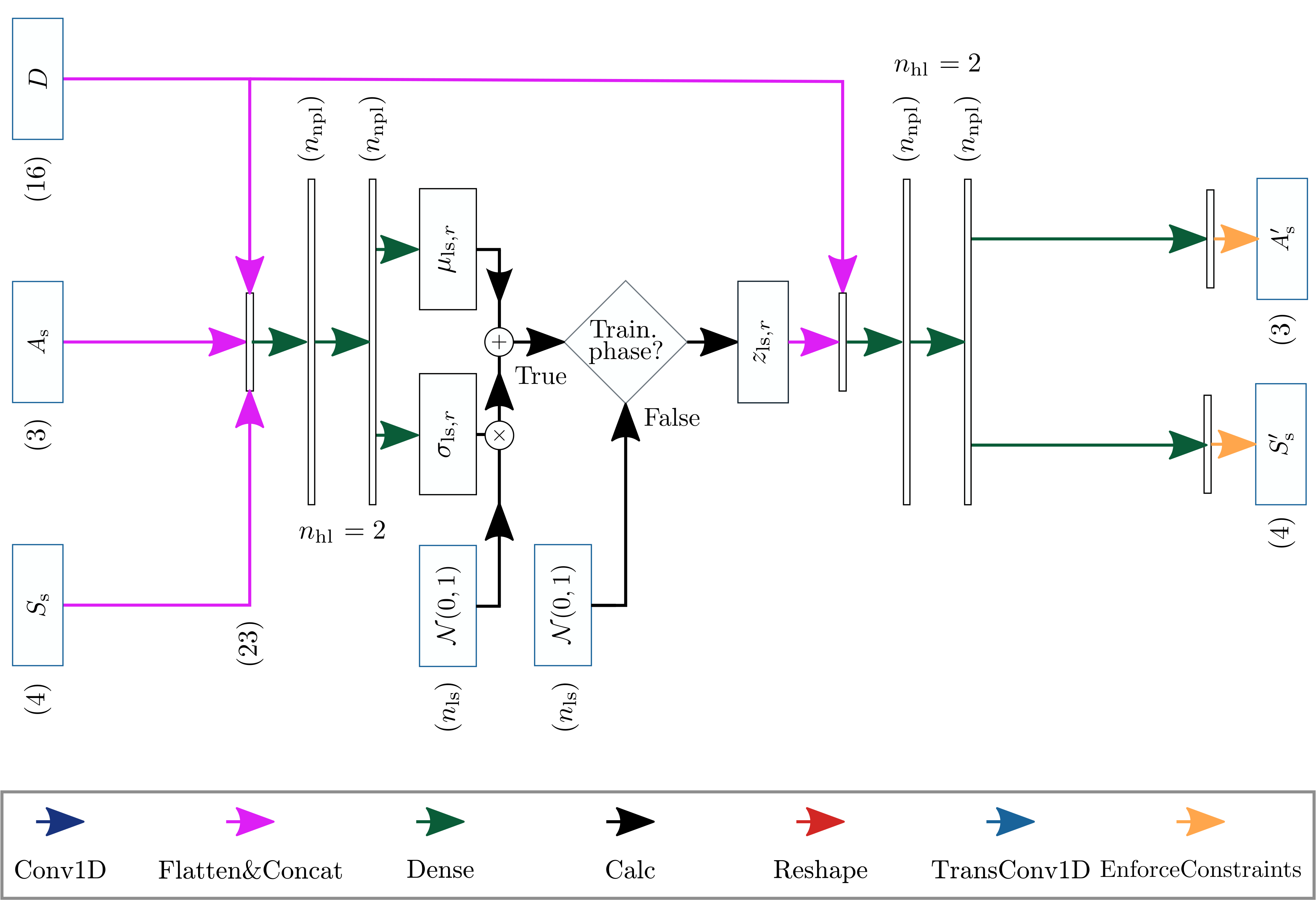}
\caption{Schematic of the DSS-CVAE network structure. The shape of the data is provided in parenthesis. Machine learning operations are indicated by colored arrows. Specifications of the convolutional operations are presented atop the arrows (i.e., kernel size $\times$ number of filters $n_\mathrm{f}$, stride).}
\label{fig:CVAE2}
\end{figure*}

%%%%%%%%%%%%%%%%%%%%%%%%%%%%%%%%%%%%%%%%%%%%%%%%%%%
\clearpage
\bibliography{references}

\end{document}